\pdfoutput=1
\documentclass[journal]{IEEEtran}
\usepackage{times}
\usepackage{cite}
\usepackage{amsmath,amssymb,amsfonts}
\usepackage{graphicx}
\usepackage{textcomp}
\usepackage{array}
\usepackage{booktabs}
\usepackage{longtable}
\usepackage{enumitem}
\usepackage{listings}
\usepackage{placeins}
\graphicspath{{figs/}}

\usepackage[hidelinks]{hyperref}

\newcommand{\orcidlink}[1]{\,\href{https://orcid.org/#1}{\raisebox{-0.15ex}{\includegraphics[height=0.92em]{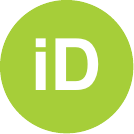}}}}
\let\PARstart\IEEEPARstart
\newcommand{\revI}[1]{#1}
\newcommand{\revblock}[1]{#1}
\newcommand{\revH}[1]{#1}

\begin{document}
\title{LLM-Based Test Oracles: Source-of-Authority Taxonomy---A Systematic Literature Review}

\author{Ali~Hassaan~Mughal\,\orcidlink{0000-0002-0724-9197},~\IEEEmembership{Graduate~Student~Member,~IEEE,}
and~Muhammad~Bilal\,\orcidlink{0000-0003-4106-0256}%
\thanks{This article has been accepted for publication in IEEE Access. This is the author's version which has not been fully edited and content may change prior to final publication. Citation information: DOI 10.1109/ACCESS.2026.3729738.}%
\thanks{This work received no external funding. The authors declare no conflicts of interest.}%
\thanks{A.~H.~Mughal is an Independent Researcher, USA (e-mail: alihassaanmughal.work@gmail.com).}%
\thanks{M.~Bilal is with the Technical University of Munich, Germany (e-mail: m.bilal@tum.de). He is the corresponding author.}}

\markboth{Accepted for publication in IEEE Access. DOI 10.1109/ACCESS.2026.3729738}%
{Mughal \MakeLowercase{\textit{et al.}}: LLM-Based Test Oracles: A Source-of-Authority Taxonomy}

\maketitle

\begin{abstract}
Large language models (LLMs) increasingly decide whether software behaves correctly, either by writing a test oracle or by acting as one. Yet two oracles can look identical and rest on different ground: one assertion encodes a written specification, another only what the model learned in training. Prior secondary studies sort oracles by form or by technique, rarely by the property that governs how far a verdict can be trusted: where its authority comes from. This systematic literature review, reported under the Preferred Reporting Items for Systematic Reviews and Meta-Analyses (PRISMA) 2020 guidelines, screens 2,436 records to 54 included studies, extended by citation searching (snowballing) to 83 in total. We read the corpus along three axes: the source of an oracle's authority, the form it takes, and the mechanism that adjudicates it. Just over half of the corpus reaches a verdict with no specification at all. That is what lets these oracles work on code with no specification to consult, and what leaves a challenged verdict with less to fall back on. Source and mechanism cross-cut rather than coincide, so a label such as LLM-as-a-judge names how a verdict is produced, not why it should be trusted. Oracle quality is most often judged by resemblance to a known oracle rather than by whether injected faults are caught. The first question to ask of any LLM oracle is therefore what one would point to in defending its verdict. The protocol, search query, and per-study coding sheet are released.
\end{abstract}


\IEEEpeerreviewmaketitle

\section{Introduction}
\PARstart{S}{oftware} now runs in almost every part of daily life, and how much we can trust it depends largely on how well we test it. A test feeds an input to a program, runs it, and then asks one question: is the result correct? The part of the test that answers that question is the \emph{test oracle}. A test is only as good as its oracle.

Deciding what counts as correct is hard. For many programs there is no easy way to know the right answer in advance. This difficulty has a name, the \emph{oracle problem}. Barr et al.~\cite{barr2015} surveyed it before large language models (LLMs) were in wide use. They organized the known ways to obtain an oracle and showed how often a strong oracle is simply missing.

LLMs have changed the picture. An LLM can now write an oracle for a program~\cite{p0259,p0098,p1131}, or act as the oracle itself by judging the output directly~\cite{p0372,p0417,p1725}. This is a fast-growing area~\cite{rW1,rH1}: most of the reviewed studies appeared in 2025 or later.

The literature describes these new oracles. It characterizes them, however, by a feature that does not, by itself, capture how far to trust the verdict. Papers sort oracles by their shape, such as an assertion~\cite{p0098} or a metamorphic relation~\cite{p0248}, or by the technique used, such as ``LLM as a judge''~\cite{p0372,p1725}. They rarely ask where the verdict's authority comes from, yet that is what sets the ceiling on how much to trust it. Two oracles can look identical yet rest on different ground. One assertion may encode a written specification; another, with the same syntax, may encode only what the model guessed.

\begin{figure*}[t]
\centering
\includegraphics[width=\textwidth]{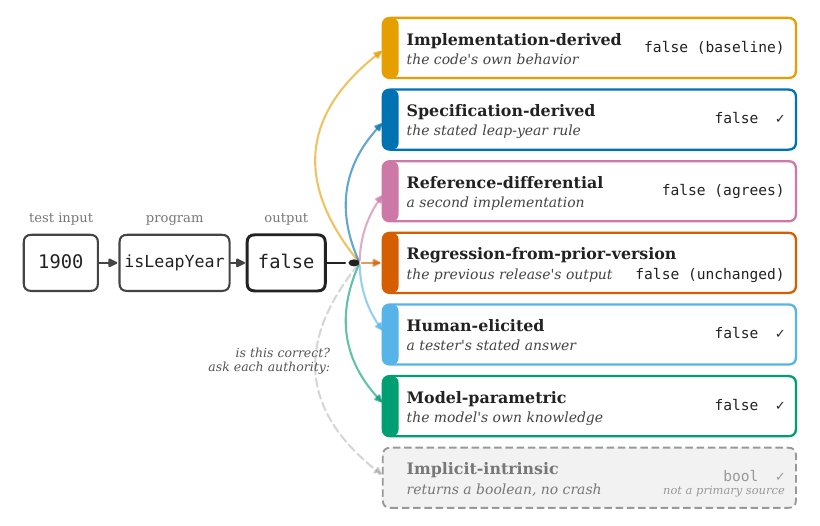}
\caption{The seven sources of oracle authority, illustrated on \texttt{isLeapYear(1900)}. \revI{The grey dashed box marks the one category no included study uses as a \emph{primary} source (it appears as a secondary source in four systems).}}
\label{fig:oracle}
\end{figure*}

This review organizes the literature by that missing feature: the \emph{source of authority}, where the correctness of the verdict ultimately comes from. \revI{We make three contributions:}
\begin{itemize}[leftmargin=1.3em, itemsep=2pt, topsep=3pt]
\item \revI{a taxonomy of oracle sources that extends Barr et al.~into the LLM era, kept separate from two questions the literature mixes in: the \emph{form} of the oracle and the \emph{mechanism} that computes pass or fail;}
\item \revI{a systematic literature review (SLR), conducted and reported under the Preferred Reporting Items for Systematic Reviews and Meta-Analyses (PRISMA)~2020 guidelines\mbox{~\cite{page2021prisma}}, that screens 2,436 records down to 54 included studies}\revI{, extended by a citation-searching (snowballing) round that adds 29 further studies, for 83 in total}\revI{;}
\item \revI{a synthesis of what the lens reveals: specification independence follows almost entirely from the source; source and mechanism cross-cut rather than coincide; }\revI{one source category goes unoccupied as a primary source of LLM-supplied oracles; }\revI{and the most frequently reported risks are weak oracles and model hallucination.}
\end{itemize}

\section{Background: The Oracle Problem in Plain Terms}
\label{sec:background}

\subsection{What a test oracle is}
Every test has four parts. There is an \emph{input}; the \emph{program under test}, often called the system under test; the \emph{output} the program produces; and the \emph{oracle}, the part that decides whether the output is correct (Fig.~\ref{fig:oracle}). The first three parts are usually easy to set up. The oracle is the difficult one.

We will use one small function as a running example through the whole article: \texttt{isLeapYear}, which should return true when a year is a leap year. A year is a leap year if it divides by 4. The exception is years that divide by 100, which are not leap years, unless they also divide by 400. So 2024 is a leap year, 1900 is not, and 2000 is. To test the function we can run it on the input 1900. Suppose it returns false. Is that correct? Answering that is the oracle's job. To answer it, the oracle must know from somewhere that 1900 is not a leap year.

\subsection{Why it is hard}
That ``from somewhere'' is the core difficulty. For \texttt{isLeapYear} the right answers are known, so an oracle is easy. For most real programs they are not known in advance and are costly to work out by hand. This is the \emph{oracle problem}: the difficulty of obtaining a reliable way to judge whether an output is correct.

Over the years, testers have found partial answers that avoid needing the exact right answer for every input. An \emph{assertion}~\cite{taromirad2025assertions} states a condition the output must meet, such as that a year reported as a leap year is divisible by 4. A \emph{metamorphic relation}~\cite{chen2018metamorphic} compares two related runs instead of checking one output against a known value. For example, running the function twice on the same year should give the same result. \emph{Differential testing}~\cite{gulzar2019differential} runs two programs that should agree and treats any disagreement as a possible bug. Each of these gives a verdict without a full description of correct behavior. Barr et al.~\cite{barr2015} surveyed these and other techniques in detail.

\subsection{What LLMs add}
An LLM can take part in the oracle in two ways. It can \emph{generate} an oracle, for example by reading a function's documentation and writing the assertions that check it~\cite{p1131,p0339,p0698}. Or it can \emph{be} the oracle, by looking at an output and judging it directly, the way a human reviewer would~\cite{p0287,p0041,p0372}. Both were impractical at scale before LLMs, and both are now common. This raises a new question, the subject of this review: when an LLM writes the oracle, or makes the verdict itself, where does the authority for that verdict come from?

\section{The Source-of-Authority Framework}
\label{sec:framework}
The central idea of our framework is to separate three questions that the literature usually treats as one.

\subsection{Three distinct questions}
When an LLM-based oracle gives a verdict, we can ask three things about it.
\begin{enumerate}[leftmargin=1.6em, itemsep=1pt]
\item \textbf{Source of authority.} Where does the correctness of the verdict ultimately come from? \revI{By \emph{authority} we mean the artifact or agent whose word makes the verdict count as correct: what one would point to when defending a pass or a fail.}
\item \textbf{Form.} What shape does the oracle take, for example an assertion or a metamorphic relation?
\item \textbf{Mechanism.} How is the pass-or-fail decision computed, for example by running an assertion or by asking the model to judge?
\end{enumerate}
Most prior work answers only the second or third question. The first sets the ceiling on trust: an oracle is only as trustworthy as the source its authority rests on. \revI{Trust here means evidential defensibility, the independent ground a verdict can appeal to, rather than a measured error rate; the ordering of sources is a conceptual claim, not an empirical result of this review. The ceiling is also set by the source's kind, not the quality of the individual artifact: a defective specification can in practice rank below a well-validated reference implementation.}

\subsection{Seven sources of authority}
The framework distinguishes seven sources of authority (Fig.~\ref{fig:oracle}). The same program output can be judged by any of them, each drawing on a different authority.

In words: an \emph{implementation-derived} oracle takes the code, or the tests already written for it, as the authority. It accepts what the code currently does, so it catches later changes but not a bug that is already present. A \emph{specification-derived} oracle takes an external description of intended behavior as the authority. The description can be formal, such as a logical postcondition; semi-formal, such as a structured rule or a Gherkin scenario; or natural-language, such as a docstring or a requirement. A \emph{reference-differential} oracle takes a separate implementation as the authority and flags disagreement. A recent variant uses several implementations the LLM writes itself and treats their agreement as the expected answer~\cite{p0699,p1985}. \revI{The strength of this source rests on how independent the reference is: an independently developed reference implementation, a differential comparison among existing implementations, and a consensus among implementations the same model generates carry successively weaker independence assumptions.} The remaining four are a \emph{model-parametric} oracle (the model's own knowledge, with no external reference), a \emph{human-elicited} oracle (criteria a person supplies), a \emph{regression-from-prior-version} oracle (an earlier version as the reference), and an \emph{implicit-intrinsic} oracle (crashes or well-formedness), each defined in Fig.~\ref{fig:oracle}.

\revI{Table\mbox{~\ref{tab:barrmap}} locates the seven sources against Barr et al.'s oracle taxonomy\mbox{~\cite{barr2015}}: five categories are retained or renamed from established oracle classes, one is subdivided, and one, model-parametric authority, is newly introduced. What prior classifications could not express is not the individual categories but the axis itself: they organize oracles by how the artifact is obtained or expressed, not by whose authority the verdict rests on, and it is the latter that the LLM era makes load-bearing.}

\begin{table}[!ht]
\centering
\caption{\revI{The seven sources of authority against Barr et al.'s oracle taxonomy\mbox{~\cite{barr2015}}.}}
\label{tab:barrmap}
\footnotesize

\revblock{%
\begin{tabular}{@{}p{2.1cm} p{2.25cm} p{3.05cm}@{}}
\toprule
This review & Barr et al. & Relation \\
\midrule
Specification-derived & Specified oracles & Retained; widened to semi-formal and natural-language specifications \\
Implementation-derived & Derived oracles & Subdivided: restricted to artifacts of the current implementation \\
Reference-differential & Pseudo-oracles (a derived subtype) & Renamed; graded by reference independence \\
Regression-from-prior-version & Derived from previous versions & Retained \\
Implicit-intrinsic & Implicit oracles & Renamed \\
Human-elicited & Human oracle & Retained; sharpened to criteria a person supplies \\
Model-parametric & (no counterpart) & Newly introduced: the model's parametric knowledge as the authority \\
\bottomrule
\end{tabular}}%
\end{table}

\subsection{Specification dependence}
The seven sources split by one question: does the oracle need an external specification? In the idealized split, only specification-derived oracles do; the other six are defined to reach a verdict without one. Section~\ref{sec:synth} examines how closely the corpus follows this split, including the few boundary cases that cross it.

\revI{Because every LLM-based oracle is driven by a prompt, it is worth fixing what counts as an external specification here. A prompt's task instruction, such as \emph{decide whether this output is correct} or \emph{write assertions for this method}, is not a specification: it directs the oracle without describing the program's intended behavior. A specification is an explicit, auditable description of that intended behavior, such as a docstring, a requirement, an application programming interface (API) contract, or a structured rule; when the verdict rests on such an artifact the oracle is specification-derived, regardless of the mechanism that consumes it and regardless of whether the artifact was supplied to the oracle or, in one boundary case, synthesized by the model itself into an executable relation (Section\mbox{~\ref{sec:synth}}). What separates this from a model-parametric oracle is auditability, not authorship: when the description of intended behavior is left implicit in the model's weights and never externalized into an inspectable artifact, the oracle is model-parametric. Whether an external specification \emph{pre-existed}, as opposed to being synthesized on the fly, is the separate axis of specification dependence. Specification dependence is therefore a property of what the oracle is given and produces, not of whether a prompt is present. The channel through which a specification reaches the oracle, whether a system instruction, the user turn, retrieved context, or a file the oracle reads, likewise does not change the coding: what matters is that an auditable description of intended behavior is present, not how it arrived.}

\revI{Two further boundaries need fixing, because the same artifact can be described from either side. Human-elicited against specification-derived turns on the artifact the verdict appeals to, not on who stands behind it: a durable, inspectable description of intended behavior is a specification whoever wrote it, while a criterion a person supplies for this judgment alone, such as a rubric written for the run or a reviewer's own decision, is human-elicited. Implementation-derived against specification-derived turns on how an existing test suite is used. We code existing tests as implementation-derived, because the reviewed systems read them as a record of what the code currently does, so the oracle inherits any fault the tests already encode. That is a coding rule and not a claim that tests never carry independent requirements: where a study rests its tests on an external requirement the coding records a secondary specification-derived source, as it does for two of the implementation-derived studies\mbox{~\cite{p0406,p2274}}. Each rule is applied through the anchoring quotation recorded for the study, so a reader who would draw the line elsewhere can see which sentence the coding rests on.}

\subsection{Source is not mechanism}
Source and mechanism cross-cut: a single source can be checked by many mechanisms, and a single mechanism can serve many sources. A specification-derived oracle, for example, can be checked by running an assertion, by matching against an expected value, or by asking the model to judge. ``LLM as a judge'' is a mechanism, not a source: the judge may be leaning on a specification, on the model's own knowledge, or on criteria a person gave it. A few mechanisms recur across different sources (Fig.~\ref{fig:alluvial}).

\section{Review Method}
\label{sec:method}

This \revI{SLR} was planned, conducted, and reported under the PRISMA 2020 statement~\cite{page2021prisma}. The protocol, the frozen search query, the operational criteria, and the per-record screening log are released as supplementary material.

\subsection{Objective}
The review maps and analyzes the literature on LLM-based test oracles according to the \emph{source} from which each oracle draws its verdict authority. A study is within scope if it (i) produces a test oracle, or (ii) empirically studies how the oracle is determined, within a pipeline that uses an LLM. Pre-2021 foundational work is retained as background through citation only and is excluded from the primary mapping counts. \revI{The design is aligned with the established methodology for SLRs in software engineering of Kitchenham and Charters\mbox{~\cite{kitchenham}} in its use of a pre-specified protocol, criterion-based screening, and structured extraction; those guidelines informed how the review was structured rather than being implemented in full. PRISMA~2020\mbox{~\cite{page2021prisma}} supplies the reporting standard.}

\subsection{Research questions}
\label{sec:rqs}
\revI{The review is organized around three research questions (RQ1--RQ3): RQ1 maps the landscape of the literature, RQ2 characterizes the corpus along the three-axis taxonomy at the review's core, and RQ3 covers how the oracles are used and where they fall short. Within RQ2, the parts on the source of authority (RQ2.a) and the adjudication mechanism (RQ2.c) are the main focus.}

\noindent\revI{\textit{RQ1 --- The landscape.}}
\begin{description}[style=unboxed, leftmargin=0pt, itemsep=4pt, topsep=3pt]
  \item[RQ1 (Landscape):] What software domains, systems under test, languages, models, and adaptation strategies characterize the LLM-based oracle literature?
\end{description}

\noindent\revI{\textit{RQ2 --- The three-axis taxonomy.} How does the corpus distribute across the source of the verdict's authority, the form the oracle takes, and the mechanism that adjudicates it? The three parts track the axes the framework keeps apart.}
\begin{description}[style=unboxed, leftmargin=0pt, itemsep=4pt, topsep=3pt]
  \item[\revI{RQ2.a} (Source of authority):] From what source does each oracle draw the authority for its verdict\revI{, and how closely does the corpus follow the specification-dependence split those source definitions imply?}
  \item[\revI{RQ2.b} (Form):] In what form is the oracle expressed (assertion, expected output value, metamorphic relation, invariant, property, classifier label, or exception oracle)?
  \item[\revI{RQ2.c} (Adjudication mechanism and LLM role):] How is the pass/fail computed, and what role does the LLM play in that computation?
\end{description}

\noindent\revI{\textit{RQ3 --- Use and gaps.} How are the oracles evaluated, how do they fail, and which regions of the taxonomy remain sparse or empty?}
\begin{description}[style=unboxed, leftmargin=0pt, itemsep=4pt, topsep=3pt]
  \item[\revI{RQ3.a} (Evaluation):] How is each oracle evaluated, and how is the oracle's own correctness assessed?
  \item[\revI{RQ3.b} (Failure modes):] What failure modes are reported, and how are they mitigated?
  \item[\revI{RQ3.c} (Gaps):] Which regions of the source-by-mechanism space are sparse or empty?
\end{description}
A guiding hypothesis, \revI{examined} in the synthesis, is that the source of authority (\revI{RQ2.a}) \revI{cross-cuts} the adjudication mechanism (\revI{RQ2.c})\revI{: the two are conceptually distinct and not in one-to-one correspondence, rather than statistically independent}. The same authority can be checked by different mechanisms, and a single mechanism such as LLM-as-judge can draw on different authorities.

\subsection{Protocol and registration}
The search query was frozen on 31~May~2026, and the inclusion and exclusion criteria were fixed in writing before screening began. In place of registration in a public registry, this pre-specified protocol is archived openly in the replication package~\cite{zenodo}.

\subsection{Eligibility criteria}
A record was \emph{included} only if all of the following held:
\begin{description}[style=unboxed, leftmargin=0pt, itemsep=4pt]
  \item[I1 (Oracle focus):] The work concerns a test oracle, that is, the mechanism that decides expected behavior or pass/fail (assertions, expected outputs, properties or invariants, metamorphic relations, differential comparison, or human judgment), rather than only test-input generation or coverage.
  \item[I2 (LLM involvement):] An LLM generates, infers, or determines the oracle, or the study empirically examines how the oracle is determined or behaves in an LLM-based testing pipeline.
  \item[I3 (Window):] Published between January~2021 and May~2026\revI{; for the citation-searching round, the window extends to that round's execution date (August~2026). Each arm's window ends on the date that arm was run. Forward citation searching necessarily reaches work published after the database search was frozen, and holding that round to the earlier cut-off would discard the studies it was added to surface. Criterion~I4 still applies, so these additions are peer reviewed like the rest of the corpus.}
  \item[I4 (Form):] At least four pages, written in English, with retrievable full text, peer reviewed and indexed (conference, journal, or early access).
\end{description}
A record was \emph{excluded} if any of the following held:
\begin{description}[style=unboxed, leftmargin=0pt, itemsep=4pt]
  \item[E1:] Testing \emph{of} LLMs or natural language processing (NLP) systems (the LLM is the system under test, for example bias, hallucination, jailbreak, or robustness testing), rather than LLMs used \emph{for} testing software.
  \item[E2:] LLM test or code generation that does not foreground an oracle (input-only, coverage-only, prefix-only, or program-repair-only).
  \item[E3:] Hardware-verification assertion or oracle generation, for example in SystemVerilog, register-transfer level (RTL), or very large-scale integration (VLSI). This is noted as an adjacent line of work in Section~\ref{sec:threats}.
  \item[E4:] Secondary studies (surveys, systematic reviews, mapping studies). These are excluded from the primary set to avoid double counting, and are retained separately for the discussion of related reviews.
  \item[E5:] Non-peer-reviewed preprints. These are deliberately excluded to keep peer review as an evidence-quality bar (criterion~I4). The resulting under-coverage of preprint venues is reported as a threat to validity (Section~\ref{sec:threats}).
  \item[E6:] Editorials, keynotes, posters, tool demonstrations, tutorials, extended abstracts, items under four pages, non-English items, and non-retrievable items.
\end{description}

\subsection{Information sources}
We searched three bibliographic databases under institutional subscription: Scopus, IEEE~Xplore, and the ACM Digital Library. All searches were executed on 31~May~2026 over the field set title, abstract, and keywords, restricted to the publication window 2021--2026. Scopus indexes a large fraction of IEEE and ACM content and therefore also serves as a cross-database consolidation source.

\subsection{Search strategy}
The search combines a \emph{population} concept (LLM terms) with a \emph{task} concept (oracle and testing terms) using a Boolean \texttt{AND}. The frozen query is:
\begin{lstlisting}
Concept A (LLM):
"large language model*" OR LLM OR LLMs OR GPT OR ChatGPT OR Codex OR Copilot
OR Llama OR "Code Llama" OR StarCoder OR Claude OR Gemini OR PaLM
OR "generative AI" OR "foundation model*" OR "code language model"

Concept B (oracle / test):
"test oracle*" OR "oracle generation" OR "oracle inference" OR "oracle problem"
OR "assertion generation" OR "test assertion*" OR "exception oracle"
OR "metamorphic testing" OR "metamorphic relation*" OR "property-based testing"
OR "differential testing" OR postcondition* OR precondition*
OR "user acceptance test*" OR "acceptance test*" OR Gherkin
OR "behavior-driven" OR "behaviour-driven" OR "behavior-driven development"

Search = Concept A AND Concept B
\end{lstlisting}
The query was rendered per database with identical Boolean logic and database-specific syntax, restricted to the computing subject area, without dropping any term, to remove cross-domain false positives such as the food-science and botany senses of ``Codex'' or ``Gherkin''. The exact per-database queries and the export procedure are in the supplementary material.

\subsection{Selection process}
\label{sec:selproc}
Records exported from the three databases were merged and de-duplicated by normalized DOI, falling back to a normalized title when a DOI was absent. This left 2,245 unique records. Because the high-recall query admitted mostly off-topic records, the corpus was too large to screen every record twice by hand. We therefore screened in two stages, an arrangement disclosed in accordance with PRISMA~2020 item~8 on the use of automation tools. The first stage was a recall-oriented automated pre-filter that removes clearly out-of-scope records; the second was independent human dual screening that verifies the retained set in full and validates the pre-filter's exclusions on a stratified sample.

\paragraph{Stage 1: automated pre-filter} An LLM (Claude Opus~4.8)\mbox{~\cite{claude48}} classified every one of the 2,245 unique records against the eligibility criteria from title and abstract. It assigned each an include, maybe, or exclude decision and, where applicable, an exclusion reason code (E1--E6). Ambiguous records were assigned \mbox{\emph{maybe}} and retained, resolving uncertainty toward full-text assessment. This stage retained 178 records and excluded 2,067. \revI{Each record was supplied to the model as its title and abstract together with the pre-registered inclusion and exclusion criteria, and the model returned an include, maybe, or exclude decision with an exclusion reason code (E1--E6), retaining any record it could not confidently exclude. Classification was run in June~2026 with the model's reasoning effort set to its maximum level. The provider's interface for this model does not expose sampling parameters such as temperature or top-p, so no decoding parameters were varied across records. The screening instruction applied criteria I1--I4 and E1--E6 to the title and abstract alone; the per-record decisions and reason codes, together with the exact screening instruction, are released for audit\mbox{~\cite{zenodo}}, and the recall and false-omission validation reported below (Table~\mbox{\ref{tab:kappa}}) is the confusion-matrix summary of this stage.}

\paragraph{Stage 2: independent human dual screening with validation} Two authors then screened a 385-record verification set, independently and blind both to the model's labels and to each other. The set comprised \mbox{\emph{all}} 178 model-retained records plus a fixed-seed, reason-stratified 10\% random sample of 207 records from the 2,067 model-excluded records. Screening all retained records catches every false positive; the exclusion sample estimates how often the model wrongly discards relevant work. Each reviewer recorded an Include, Maybe, or Exclude decision with a one-line rationale for every record.

Inter-rater agreement was substantial, measured by Cohen's \mbox{$\kappa$}\mbox{\revI{\mbox{~\cite{cohen1960}}}}, a chance-corrected measure of how often two reviewers agree. On the operative retain-versus-exclude decision (Retain \mbox{$=$} Include \mbox{$\cup$} Maybe), \mbox{$\kappa = 0.79$} (95\% bootstrap\mbox{\revI{\mbox{~\cite{efron}}}} confidence interval (CI) \mbox{$[0.72, 0.85]$}; 90.4\% raw agreement). On the finer three-level scale it was \mbox{$\kappa = 0.64$} (95\% CI \mbox{$[0.57, 0.71]$}; 81.6\% agreement). The 71 disagreements were reconciled conservatively. Include-versus-Maybe splits, where both reviewers leaned toward retention, were retained. The eight direct Include-versus-Exclude conflicts were decided individually against the full text. Exclude-versus-Maybe splits, where neither reviewer favored inclusion, were excluded. The retain-on-uncertainty principle stated above governs the recall-oriented pre-filter; this human stage is where precision is applied. Reconciliation yielded a consensus of 115 retained records (87 include, 28 maybe) and 270 excluded over the verification set.

\paragraph{Validation of the pre-filter} Treating the human consensus as the reference, the model achieved 99.1\% recall on the verification set, agreeing with all but one consensus-retained record. That single miss was itself a borderline case reconciled to include only after full-text discussion. On the stratified sample of its own exclusions, the model's observed false-omission rate was \mbox{$1/207 = 0.48\%$} \revI{(exact binomial, or Clopper--Pearson\mbox{~\cite{clopperpearson}}, 95\% CI \mbox{$0.01\%$}--\mbox{$2.66\%$})}. If this rate held across the 1,860 model-excluded records not individually re-screened, it would imply roughly nine missed relevant records \revI{(the same interval scaled to the 1{,}860 unaudited exclusions gives roughly \mbox{$0$} to \mbox{$50$})}. Because the CI around a single observed omission is wide, this residual risk is carried as a threat to validity (Section~\mbox{\ref{sec:threats}}). The sampled exclusions and their decisions are also released for audit\mbox{~\cite{zenodo}}. The model was otherwise over-inclusive, retaining 64 consensus-excluded records. Its precision is supplied by the human stage. \revI{Table~\mbox{\ref{tab:confusion}} sets out the confusion matrix these figures summarize.}

\subsection{Data extraction and coding}
\label{sec:codebook}
Each included study is coded against a 35-field codebook, fixed before extraction and released as supplementary material. The controlled vocabularies are likewise fixed in advance. The codebook is built on the three axes of Section~\ref{sec:framework}: the source of authority (\revI{RQ2.a}), the form (\revI{RQ2.b}), and the adjudication mechanism (\revI{RQ2.c}). It also adds descriptive fields for the landscape, the evaluation, and the reported failure modes (RQ1, \revI{RQ3.a}, \revI{RQ3.b}). Source and mechanism are coded \revI{as two separate variables} (Section~\ref{sec:framework}). A primary source and, where a system is compositional, a secondary source are recorded.

\revI{All 115 retained records, including the one later found to duplicate another, were coded against the locked codebook by the authors and then re-assessed in a second pass covering the two variables the review's conclusions rest on, eligibility and the primary source of authority, each supported by its own quotation drawn from the full text. The passes concurred on 80.7\% of eligibility decisions (22 differences over the 114 unique full-text records, the duplicated pair counting once) and on 83.3\% of primary-source codes (8 differences over the 48 records both passes judged includable; the remaining included studies were admitted only at adjudication, so the second pass assigned them no source). The 29 studies added by the citation-searching round were coded in two full passes over all fields, the second revising the primary source in four of the 29, a concordance of 86.2\% (Table~\mbox{\ref{tab:codeconf}}). Two of those four revisions moved a study to model-parametric authority, one from implementation-derived and one from human-elicited, which is where the categories are hardest to hold apart in practice and where the boundary rules of Section~\mbox{\ref{sec:framework}} do most of their work. Every difference was adjudicated against the pre-registered boundary rules, the four hardest by a fresh full-text re-read; six eligibility decisions and two primary-source codes were resolved in favor of the first pass, and the two contested source codings\mbox{~\cite{p0184,p1850}} are treated as boundary cases in Section~\mbox{\ref{sec:threats}}. Both coding passes and the per-record adjudication decisions are released\mbox{~\cite{zenodo}}, so the differences can be inspected or re-quantified independently. Each study's coding carries an anchoring quotation from its full text, so any coding can be re-traced and contested from the released per-study sheet.} The synthesis is descriptive: it reports distributions across the taxonomy and reads gaps off the sparse or empty regions of the source-by-mechanism space. Effectiveness numbers are not pooled, because the studies differ too much in benchmark and metric: \revI{64\%} use a custom benchmark. A lightweight, non-exclusionary quality appraisal of the included studies, scored on \revI{nine} items derived from these coded fields, is reported in Section~\ref{sec:quality}.

\begin{table}[!ht]
\centering
\caption{\revI{Where the two coding passes placed each study's primary source of authority, for the citation-searching arm. Rows are the first pass, columns the second; the diagonal is agreement. The same matrix cannot be shown for the database arm, whose first pass was revised in place when the adjudication was applied. Column abbreviations follow Table\mbox{~\ref{tab:appendix}}.}}
\label{tab:codeconf}
\setlength{\tabcolsep}{3.5pt}\footnotesize
\revblock{%
\begin{tabular}{@{}l ccccccr@{}}
\toprule
1st\,$\downarrow$ / 2nd\,$\rightarrow$ & Human & Impl & Model & Ref-diff & Regr & Spec & Total\\
\midrule
Human    & 5 & . & 1 & . & . & . & 6\\
Impl     & . & 6 & 1 & . & . & . & 7\\
Model    & . & . & 6 & . & . & . & 6\\
Ref-diff & . & . & . & . & 1 & . & 1\\
Regr     & . & . & . & . & 1 & . & 1\\
Spec     & . & . & . & 1 & . & 7 & 8\\
\midrule
Total    & 5 & 6 & 8 & 1 & 2 & 7 & 29\\
\bottomrule
\end{tabular}}%
\end{table}

\subsection{Selection results}
\label{sec:flow}
\begin{figure*}[tb]
\centering
\includegraphics[width=\textwidth]{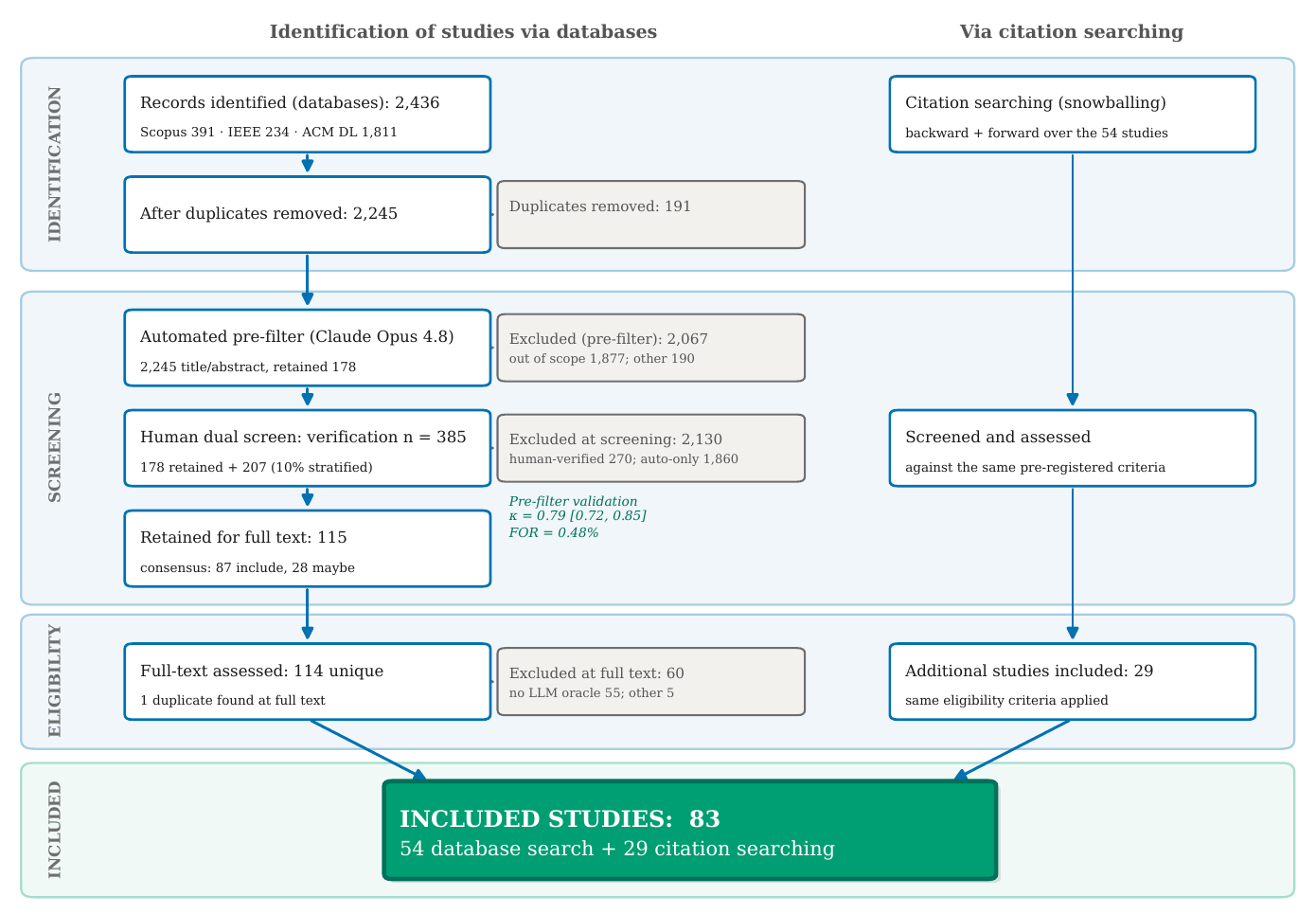}
\caption{\revI{PRISMA 2020 flow of records: the database-search arm (left, full detail) and the citation-searching arm (right, summarized), converging on the 83 included studies. Per-stage exclusion reasons are in the supplementary material.}}
\label{fig:prisma}
\end{figure*}

Figure~\ref{fig:prisma} and Table~\ref{tab:kappa} report the record flow and screening reliability. The searches identified 2,436 records, de-duplicated to 2,245. The dominant pre-filter exclusion reason was out-of-scope topical drift, records that matched the broad query terms but did not concern LLM-based software test oracles. \revI{In addition, a backward and forward citation-searching (snowballing) round over the 54 studies included from the database search, executed in August~2026 and screened against the same eligibility criteria, contributed 29 further included studies; the review therefore synthesizes 83 studies in total (Fig.\mbox{~\ref{fig:prisma}}).}

\begin{table*}[tb]
\centering
\caption{Screening and coding reliability, and pre-filter validation. The first block covers the 385-record verification set; \revI{the second reports how far the two coding passes concurred, before adjudication}.}
\label{tab:kappa}
\small
\begin{tabular}{@{}p{10.4cm} p{6.3cm}@{}}
\toprule
Measure & Value \\
\midrule
Reviewer agreement, binary (retain vs.\ exclude) & $\kappa = 0.79$; 95\% CI $[0.72, 0.85]$; 90.4\% \\
Reviewer agreement, three-level (incl./maybe/excl.) & $\kappa = 0.64$; 95\% CI $[0.57, 0.71]$; 81.6\% \\
Pre-filter recall vs.\ consensus & 99.1\% (114/115) \\
Pre-filter false-omission rate (sampled exclusions) & 0.48\% (1/207; \revI{95\% CI 0.01--2.66\%;} $\approx$\,9 projected over 1{,}860\revI{, CI $\approx$0--50}) \\
Pre-filter over-inclusion (consensus-excluded yet retained) & 64 \\
\midrule
\multicolumn{2}{@{}l}{\revI{\emph{Concordance between the two coding passes, before adjudication}}} \\
\revI{Eligibility (114 unique full-text records)} & \revI{80.7\% (92/114)} \\
\revI{Primary source (48 records both passes judged includable)} & \revI{83.3\% (40/48)} \\
\revI{Primary source, citation-searching arm (29 studies)} & \revI{86.2\% (25/29)} \\
\revI{Differences resolved in favor of the first pass} & \revI{8 (6 eligibility, 2 source)} \\
\bottomrule
\end{tabular}
\end{table*}

\begin{table}[!ht]
\centering
\caption{\revI{Pre-filter confusion matrix over the 385-record verification set: all 178 model-retained records together with a reason-stratified 10\% sample of 207 of the 2,067 model-excluded records. The exclusion column is a sample rather than a population total, so it bounds the false-omission rate but does not give a corpus-wide precision.}}
\label{tab:confusion}
\small
\revblock{%
\begin{tabular}{@{}l c c c@{}}
\toprule
 & \multicolumn{2}{c}{Human consensus} & \\
\cmidrule(lr){2-3}
Pre-filter decision & Retain & Exclude & Total \\
\midrule
Retain  & 114 & 64  & 178 \\
Exclude & 1   & 206 & 207 \\
\midrule
Total   & 115 & 270 & 385 \\
\bottomrule
\end{tabular}}%
\end{table}

\subsection{Full-text eligibility outcome}
\label{sec:ft}
The 115 retained records were assessed at full text against the locked criteria of Section~\ref{sec:method}, recording for each eligibility decision a supporting evidence quote together with the codebook fields of Section~\ref{sec:codebook}. One record was found to duplicate another, leaving 114 unique full-text records, of which 54 were included~\cite{p0001,p0009,p0014,p0017,p0018,p0041,p0043,p0048,p0072,p0083,p0087,p0091,p0098,p0103,p0129,p0132,p0137,p0161,p0177,p0183,p0184,p0211,p0213,p0223,p0231,p0248,p0259,p0261,p0279,p0287,p0304,p0316,p0320,p0323,p0339,p0371,p0372,p0388,p0390,p0406,p0408,p0413,p0417,p0464,p0698,p0699,p1131,p1154,p1311,p1725,p1846,p1850,p1985,p2138} and 60 excluded. \revI{The citation-searching round contributed 29 further included studies~\cite{p2246,p2247,p2248,p2249,p2250,p2251,p2252,p2253,p2254,p2255,p2256,p2257,p2258,p2259,p2260,p2261,p2262,p2263,p2264,p2265,p2266,p2267,p2268,p2269,p2270,p2271,p2272,p2273,p2274}, so the syntheses that follow cover 83 studies.}

The dominant exclusion reason, by a wide margin, is E2 (55 of 60). On full reading, these studies use the LLM to generate test inputs, to produce specifications that a static solver then proves, to drive exploration whose only failure signal is a crash or a log, or to generate test documents for human execution. In each, the verdict is reached by a non-LLM mechanism, a differential comparator, a static verifier, or a human reviewer, rather than by an oracle the LLM supplies or determines. This criterion cannot be applied reliably from title and abstract, which accounts for the gap between the 115 apparently oracle-relevant records and the 54 in which an LLM supplies or determines the oracle. The remaining exclusions are three items that are too short or non-research (E6) and two that fail another inclusion criterion. One boundary recurred often enough to state explicitly: an LLM-generated invariant, contract, or assertion that is \emph{executed against a running system} is treated as an oracle and included. The same artifact used only inside a \emph{static proof}, by contrast, is treated as verification and excluded.

\section{Findings}
\label{sec:synth}
Every number below is computed from the released coding sheet\revI{~\mbox{\cite{zenodo}}} and can be reproduced; each study's coding is listed in Table~\ref{tab:appendix}.

\subsection{The landscape: who, what, and when (RQ1)}
\begin{figure}[!ht]
\centering
\includegraphics[width=\columnwidth]{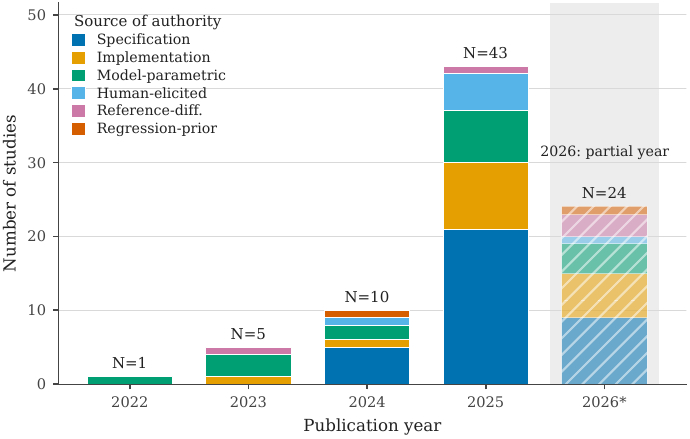}
\caption{Included studies by publication year, stacked by source of authority (2026 partial, hatched).}
\label{fig:corpus}
\end{figure}

\begin{figure}[!ht]
\centering
\includegraphics[width=0.82\columnwidth]{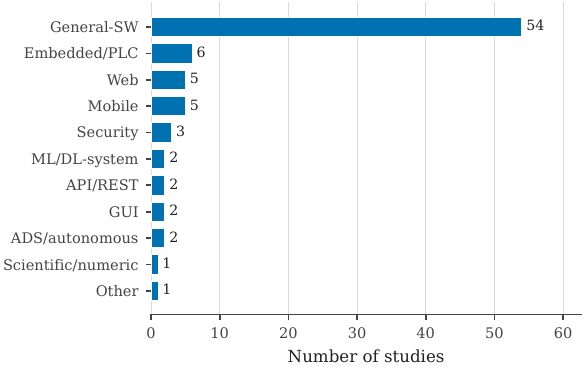}
\caption{Application domains of the systems under test. Abbreviated labels: General-SW, general-purpose software; ML/DL, machine or deep learning; PLC, programmable logic controller; API, application programming interface; REST, representational state transfer; GUI, graphical user interface; ADS, autonomous driving system.}
\label{fig:domains}
\end{figure}

The reviewed work is recent and concentrated, most of it targeting ordinary software at the unit level: general-purpose software accounts for \revI{54 of the 83 studies. The rest form smaller clusters: embedded or programmable logic controller (PLC) control with six, web and mobile with five each, security with three, two studies each in representational state transfer (REST) or API services, graphical user interfaces (GUIs), machine or deep learning (ML/DL) systems, and autonomous driving systems (ADSs), and the remaining two domains contributing one study each} (Fig.~\ref{fig:domains}). The source of authority varies with the domain. Implementation-derived oracles are almost all general-purpose unit-level software, \revI{16 of 17. Model-parametric oracles split between the neural assertion-generation line on general-purpose software (8 of 17) and interactive or visual domains such as mobile apps, GUIs, and ADSs (7 of 17)\mbox{~\cite{p0041,p0316}}, with the remaining two in ML/DL and embedded code. In the interactive domains a plausible reason is that no executable specification is on offer}, leaving the model's own expectations as often the only authority within reach. \revI{One study is from 2022, five are from 2023, ten from 2024, 43 from 2025, and 24 from 2026. As a result, 81\%} appeared in 2025 or later (Fig.~\ref{fig:corpus}). All included work is peer reviewed, spread across \revI{40} venues but led by the main software-engineering ones (Table~\ref{tab:venues}). Conferences carry most of it \revI{(51 of 83), with journals a strong second (29) and workshops the remaining three}.

\begin{table}[!ht]
\centering
\caption{Leading publication venues (those with at least two studies; \revI{40} venues in total).}
\label{tab:venues}
\small

\revblock{%
\begin{tabular}{@{}p{6.4cm} r@{}}
\toprule
Venue & Studies \\
\midrule
Proc. ACM Software Engineering (FSE / ISSTA) & 9 \\
Int. Conf. Software Engineering (ICSE, incl. SEIP) & 7 \\
ACM Trans. Software Engineering and Methodology (TOSEM) & 6 \\
IEEE Transactions on Software Engineering (TSE) & 6 \\
Int. Conf. Automated Software Engineering (ASE) & 5 \\
Int. Conf. Foundations of Software Engineering (FSE, incl. ESEC) & 3 \\
IEEE Conf. Software Testing, Verification and Validation (ICST) & 3 \\
Annual Meeting of the Assoc. for Computational Linguistics (ACL) & 3 \\
Information and Software Technology & 3 \\
Lecture Notes in Computer Science (various) & 3 \\
Int. Conf. AI-powered Software (AIware) & 2 \\
Int. Conf. Automation of Software Test (AST) & 2 \\
Asia-Pacific Software Engineering Conf. (APSEC) & 2 \\
Int. Symp. Software Testing and Analysis (ISSTA proceedings) & 2 \\
IEEE Int. Conf. Software Testing Workshops (ICSTW) & 2 \\
\midrule
\multicolumn{2}{@{}l}{\footnotesize 25 further venues with one study each.} \\
\bottomrule
\end{tabular}}%
\end{table}

\begin{figure*}[tb]
\centering
\includegraphics[width=\textwidth]{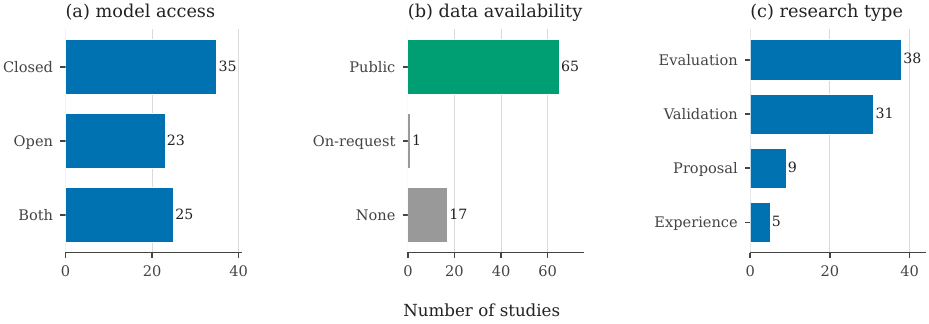}
\caption{Study characteristics: (a) model access, (b) data or code availability, (c) research type.}
\label{fig:studychars}
\end{figure*}

\begin{figure*}[tb]
\centering
\includegraphics[width=\textwidth]{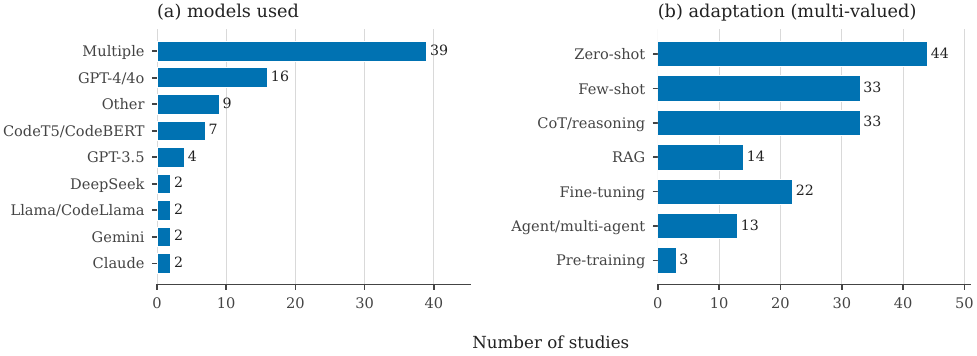}
\caption{LLM use: (a) primary model family per study, (b) adaptation strategies (multi-valued). CoT, chain-of-thought; RAG, retrieval-augmented generation.}
\label{fig:techniques}
\end{figure*}

The studies lean on closed-weight models and on prompting rather than training. Setting aside the \revI{39 studies that compare several models, the GPT family is the most common single choice: 20 studies use a GPT model as their primary, a GPT-4-class model in 16 and GPT-3.5 in 4}. Most use the model as it comes, with fine-tuning in \revI{22} (Fig.~\ref{fig:techniques}). Closed models \revI{(35)} outnumber open-weight ones \revI{(23)}\revI{, with 25 studies using both} (Fig.~\ref{fig:studychars}(a)). Open-science practice is common: \revI{65 of 83} release code or data (Fig.~\ref{fig:studychars}(b)), and most studies evaluate or validate existing ideas rather than propose new ones (Fig.~\ref{fig:studychars}(c)). The languages tested are mostly Java \revI{(35)} and Python \revI{(18)}.

\subsection{Where the authority comes from (\revH{RQ2.a})}
\begin{figure}[!ht]
\centering
\includegraphics[width=0.9\columnwidth]{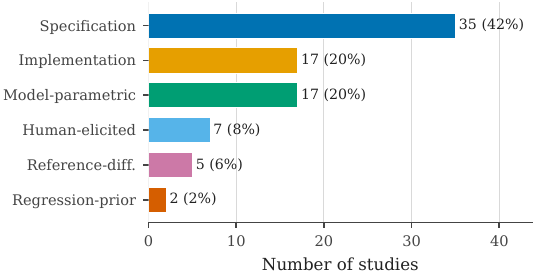}
\caption{Primary source of oracle authority (\revI{83} studies).}
\label{fig:source}
\end{figure}

\revI{The single largest group of reviewed oracles draws} its authority from a written description of intended behavior (Fig.~\ref{fig:source}). Specification-derived oracles \revI{account for 35 of the 83}. Within them, natural-language artifacts, such as docstrings, code comments, requirements, and API documentation, account for \revI{23}, ahead of semi-formal (7) and formal \revI{(5)}. \revI{Implementation-derived oracles, which read expected behavior from the code or existing tests, and model-parametric oracles, which rely on the model's own knowledge with no external reference, tie as the next largest groups (17 each). Human-elicited oracles follow (7).} Reference-differential oracles \revI{(5)} include a recent variant in which the consensus output of several model-generated alternative implementations, rather than an independent reference version, supplies the expected value~\cite{p0699,p1985}. \revI{One of the seven categories, implicit-intrinsic, remains unoccupied as a \emph{primary} source, appearing only as a secondary signal in compositional systems. Regression-from-prior-version, unoccupied in the database-search corpus, is now held by two studies in which a prior version's behavior anchors an LLM-generated or LLM-judged test\mbox{~\cite{p2260,p2271}}} (Section~\ref{sec:gaps}).

\revI{Recording one primary source per study is a coding decision rather than a property of the systems, so it is worth reading the corpus a second way, with each study's secondary source counted alongside its primary one. Sixty-one of the 83 studies record a secondary source, giving 144 source labels in all. The picture is stable where it matters most: specification-derived remains the largest group by a wide margin (42 labels), and reference-differential and regression-from-prior-version stay near the bottom (8 and 7). Two things move. Implementation-derived authority passes model-parametric (31 against 27), because code and existing tests are the most common supporting signal in compositional designs; and implicit-intrinsic, which no study uses as its primary source, appears as a secondary signal in four. The conclusions that follow rest on the primary coding and on the source-by-mechanism structure, and neither is disturbed by this second reading; what it does show is that the one empty category is a consequence of counting primary sources only, not a claim that the signal is absent from the corpus. Every secondary source is listed study by study in Table~\mbox{\ref{tab:appendix}}.}

\subsection{\revH{Specification dependence: an audit of the source coding (RQ2.a)}}
\begin{figure}[!ht]
\centering
\includegraphics[width=\columnwidth]{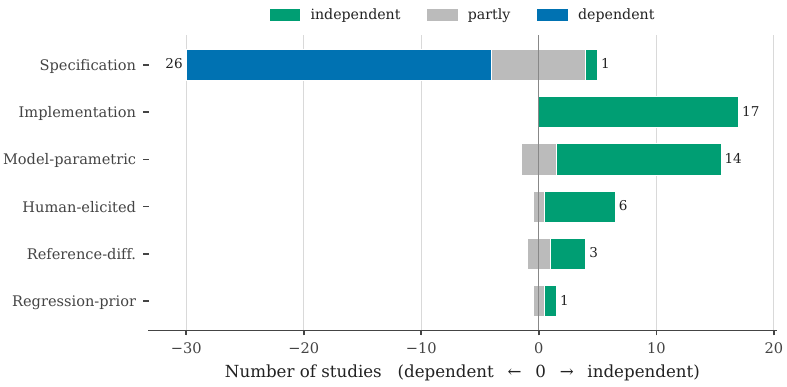}
\caption{Specification (in)dependence of the verdict, by primary source of authority.}
\label{fig:independence}
\end{figure}

\revI{Specification dependence is not posed as a separate research question, because the source categories are defined by their relation to a specification; it is reported here as an audit of the RQ2.a coding.} Because the sources are defined by their relation to a specification, the corpus follows the idealized split \revI{in direction, though with a wider partial fringe than the database-search corpus alone} (Fig.~\ref{fig:independence}). \revI{This is a consistency check rather than an empirical discovery. Specification dependence is coded as its own field, separately from the source code, so the near-perfect alignment between the two confirms that the categories behave as defined; this alignment is what is expected by construction and is reported as an audit of the taxonomy, not as a finding about the literature; the substantive result is the source distribution reported above, and the specification-independent share simply follows from it.} \revI{Every implementation-derived oracle (17 of 17) renders its verdict without any external specification, as do 14 of the 17 model-parametric, six of the seven human-elicited, three of the five reference-differential, and one of the two regression-from-prior-version oracles; the remainder are partly dependent, and no non-specification source is fully specification-dependent. Specification-derived oracles, by contrast, are specification-dependent or partly so in 34 of 35 cases. The 42 fully specification-independent studies make up just over half of the corpus. All but one draw their verdict authority from the code, the model's own knowledge, an independent reference, a prior version, or a human; the exception is the specification-derived oracle that infers its own specification, discussed next.}

\revI{The two crossings carry the real information. The one specification-derived oracle that needs no pre-existing specification\mbox{~\cite{p0261}} has the LLM \emph{infer} executable metamorphic relations; the inferred relation then serves as the specification, so the spec is synthesized rather than supplied. In the database-search corpus, the one reference-differential oracle that partly depends on a specification\mbox{~\cite{p1985}} takes the consensus of LLM-synthesized reference versions as its expected output, and its coding records a secondary specification-derived source: part of the verdict's authority flows from the specification that guided the synthesis. Both show the same pattern: the boundary blurs exactly where the LLM manufactures the artifact that other designs take as given.} \revI{The citation-searching extension adds six further partial crossings of the same character, among them the corpus's second partly-dependent reference-differential oracle: in each, part of the verdict's authority flows through a specification-shaped artifact that the model reads, synthesizes, or internalizes, from docstring signals inside neural assertion generators to the issue text behind a regression-catching test.}

\subsection{How oracles are expressed and adjudicated (\revH{RQ2.b, RQ2.c})}
Table~\ref{tab:formmech} reports oracle form and adjudication mechanism, both multi-valued, so column totals exceed \revI{83}. Assertions dominate the form \revI{(56)} and runtime assertion execution the mechanism \revI{(56)}. The key pattern is the division of labor: \revI{in 65 of 83 studies the LLM generates an oracle artifact that some other mechanism then checks, in 16 it is itself the judge, and in two it validates an oracle produced by other means.}

\begin{table}[!ht]
\centering
\caption{Oracle form, adjudication mechanism, and LLM role (form and mechanism multi-valued; totals exceed \revI{83}). PBT, property-based testing.}
\label{tab:formmech}

\revblock{%
\begin{tabular}{@{}l r@{}}
\toprule
\multicolumn{2}{@{}l}{\emph{Oracle form (\revI{RQ2.b})}} \\
\midrule
Assertion & 56 \\
Expected-output value & 19 \\
Classifier / judge label & 15 \\
Metamorphic relation & 12 \\
Invariant & 11 \\
Property / PBT & 9 \\
Other (exception oracle) & 2 \\
\midrule
\multicolumn{2}{@{}l}{\emph{Adjudication mechanism (\revI{RQ2.c})}} \\
\midrule
Runtime assertion execution & 56 \\
Exact / structural match & 26 \\
LLM-as-judge & 22 \\
Human review & 11 \\
Statistical / distributional & 8 \\
Formal verification & 6 \\
Semantic / similarity & 4 \\
\midrule
\multicolumn{2}{@{}l}{\emph{LLM role (\revI{RQ2.c})}} \\
\midrule
Generates oracle artifact (checked otherwise) & 65 \\
Is the oracle / judge & 16 \\
Validates an oracle produced by other means & 2 \\
\bottomrule
\end{tabular}}%
\end{table}

\begin{figure*}[tb]
\centering
\includegraphics[width=\textwidth]{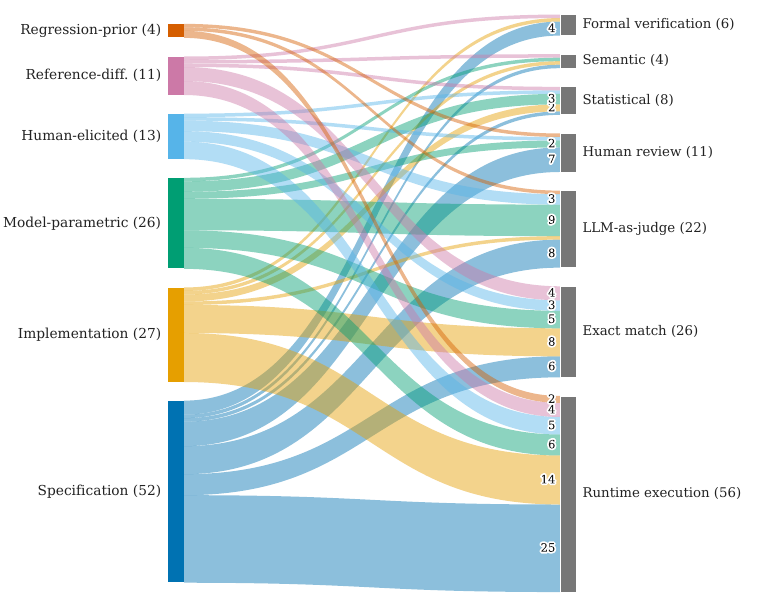}
\caption{Source of authority (left) to the mechanism that adjudicates it (right). Parenthetical numbers are node totals. Each ribbon's number is that source's contribution. All are study co-occurrences (one per mechanism a study uses), so the source totals exceed the primary-source counts of Fig.~\ref{fig:source}. Unlabeled ribbons are single studies. \revI{Each study contributes its single primary source paired with each mechanism it uses, so no source combinations are introduced; a numerical version is given in Table\mbox{~\ref{tab:srcmech}}.}}
\label{fig:alluvial}
\end{figure*}

The two axes do not line up one to one (Fig.~\ref{fig:alluvial}). \revI{Specification-derived authority is decided by all seven mechanisms; conversely each mechanism draws on between three and six different sources. Runtime assertion execution, the most common mechanism (56 co-occurrences), serves every occupied source. LLM-as-judge (22) serves specification-derived, model-parametric, human-elicited, and regression-from-prior-version authority, and appears once as a validation step over an implementation-derived oracle\mbox{~\cite{p2253}}. The LLM's role adds a second pattern. In all 16 studies where the LLM is itself the judge, its authority is a specification, its own parametric knowledge, human-elicited criteria, or a prior version, and in none is it implementation-derived or reference-differential. With only 16 such studies this is a pattern in the corpus rather than an established constraint.} For those two sources the LLM writes an artifact that some other mechanism runs. Cataloging oracles by mechanism or form alone, as prior secondary studies do, discards this structure.

\begin{table*}[!t]
\centering
\caption{\revI{Numerical source-by-mechanism co-occurrence matrix behind Fig.\mbox{~\ref{fig:alluvial}}: one count per (source, mechanism) pair a study realizes, over the seven principal mechanisms. Row and column totals are the node totals of the figure (}\revI{133 co-occurrences across the 83 studies}\revI{). Three studies additionally record an \emph{other} mechanism (mutation testing, a differential cross-check, and a rule-based matcher), listed in the appendix but outside these seven. Sources are single-valued per study, so no source combinations arise.}}
\label{tab:srcmech}
\setlength{\tabcolsep}{6pt}\footnotesize

\revblock{%
\begin{tabular}{@{}l ccccccc c@{}}
\toprule
Source & Runtime exec & Exact match & LLM-judge & Human review & Statistical & Semantic & Formal verif & Total\\
\midrule
Specification-derived & 25 & 6 & 8 & 7 & 1 & 1 & 4 & 52\\
Implementation-derived & 14 & 8 & 1 & 0 & 2 & 1 & 1 & 27\\
Model-parametric & 6 & 5 & 9 & 2 & 3 & 1 & 0 & 26\\
Human-elicited & 5 & 3 & 3 & 1 & 1 & 0 & 0 & 13\\
Reference-differential & 4 & 4 & 0 & 0 & 1 & 1 & 1 & 11\\
Regression-from-prior-version & 2 & 0 & 1 & 1 & 0 & 0 & 0 & 4\\
\midrule
Total & 56 & 26 & 22 & 11 & 8 & 4 & 6 & 133\\
\bottomrule
\end{tabular}}%
\end{table*}

\subsection{How well the oracles work (\revH{RQ3.a})}
Most studies compare against a baseline \revI{(65 of 83)}, but the evaluation ground is fragmented and the oracle's own correctness is unevenly checked. \revI{64\% rely on a custom benchmark (53 of 83). The most common shared benchmarks are Defects4J (16) and HumanEval (7)}, which limits cross-study comparison. Ground truth is most often human-provided \revI{(37)} or taken from existing tests \revI{(20)}, with reference implementations \revI{(15)} and specifications \revI{(8)} behind. Oracle quality is assessed against ground-truth oracles in \revI{52} studies and by manual inspection in \revI{19}; only \revI{13} use mutation analysis to test whether the oracle actually catches injected faults, and 3 do not assess it at all. The dominant reported metrics are accuracy against a reference \revI{(39)} and fault detection \revI{(35)}, followed by precision and recall \revI{(31)}, coverage \revI{(19)}, and mutation score \revI{(19)}. Resemblance to a known oracle is thus the dominant quality signal, rather than mutation-grounded fault detection.

\subsection{How the oracles fail (\revH{RQ3.b})}
\begin{figure}[!ht]
\centering
\includegraphics[width=0.82\columnwidth]{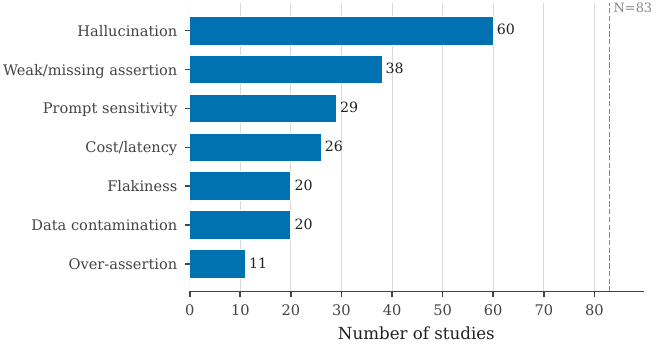}
\caption{Reported failure modes (multi-valued; dashed line marks the \revI{83} studies). One-off study-specific modes are omitted.}
\label{fig:failure}
\end{figure}

Reported failure modes converge on a small set (Fig.~\ref{fig:failure}). Hallucination is the model asserting behavior the system does not exhibit. It is the hazard studies engage with most, surfacing in \revI{60 of 83} studies. A closer reading separates three overlapping ways studies engage with it (Table~\ref{tab:halluc}): \revI{58} name it as a risk, \revI{44} report observing it empirically in their own results, and \revI{46} apply a concrete mitigation. Among the other modes, the leaders are \revI{weak or missing assertions (38), where the generated oracle is too loose to detect the faults it should, and prompt sensitivity (29); then cost or latency (26), flakiness or nondeterminism (20), and data contamination (20)}. Over-assertion, an oracle so strict that it rejects correct behavior, appears in only \revI{11} studies, against \revI{38} for weak or missing assertions, making weak oracles the most frequently reported correctness risk. But how often a failure is reported may not match how often it happens. The \revI{46} mitigations are led by execution-based filtering, compiling and running a candidate oracle and discarding what fails to compile or rejects known-correct behavior~\cite{p0698,p0388}; self-consistency or multi-sample voting~\cite{p0699,p1985}; and human review (\revI{11} studies, Table~\ref{tab:formmech}).

\begin{table}[!ht]
\centering
\caption{How studies engage with hallucination, the most common failure mode (categories overlap).}
\label{tab:halluc}

\revblock{%
\begin{tabular}{@{}l r@{}}
\toprule
Engagement with hallucination & Studies \\
\midrule
Named as a risk or threat & 58 \\
Empirically observed in own results & 44 \\
Mitigation applied & 46 \\
\midrule
\multicolumn{2}{@{}l}{\footnotesize 60 of 83 studies engage in at least one of these ways.} \\
\bottomrule
\end{tabular}}%
\end{table}

\subsection{Study quality appraisal}
\label{sec:quality}
To characterize the rigor of the corpus without excluding any study, each of the \revI{83} included studies was scored on a lightweight \revI{nine-item} checklist derived from the coded fields (Table~\ref{tab:quality}). The corpus is uniformly strong on the items the taxonomy relies on: every study defines its oracle, identifies an authority source, and discusses its own limitations. Almost all also report a benchmark \revI{(75 of 83)} and a ground-truth source \revI{(80)}. It is weakest exactly where the evaluation findings (\revI{RQ3.a}) already pointed: fault detection is evaluated in \revI{35} studies and \revI{mutation analysis used, as a metric or as an oracle check, in only 20}. \revI{We report the nine items individually rather than as a summed score: they mix reporting, artifact-availability, and fault-detection constructs that do not form a single validated quality index. Read item by item, they describe} a literature that is mature in reporting but light on fault-grounded evaluation\revI{, so findings that lean chiefly on the fault-grounded items (RQ3.a) rest on the weaker part of the corpus}.

\begin{table}[!ht]
\centering
\caption{Lightweight quality appraisal of the \revI{83} included studies (non-exclusionary; each item scored 0/1).}
\label{tab:quality}

\revblock{%
\begin{tabular}{@{}l r@{}}
\toprule
Quality item & Studies \\
\midrule
Oracle explicitly defined & 83 (100\%) \\
Authority source identifiable & 83 (100\%) \\
Threats / limitations discussed & 83 (100\%) \\
Ground truth available & 80 (96\%) \\
Dataset / benchmark described & 75 (90\%) \\
Code or data released & 65 (78\%) \\
\revI{Baseline comparison reported} & \revI{65 (78\%)} \\
Fault detection evaluated & 35 (42\%) \\
Mutation analysis used (metric or oracle check) & 20 (24\%) \\
\midrule
\multicolumn{2}{@{}l}{\footnotesize \revI{Items reported individually; not summed into a composite score.}} \\
\bottomrule
\end{tabular}}%
\end{table}

\section{Discussion}
\label{sec:discussion}

\subsection{What the findings mean}
Three results stand out, together showing a field that has changed what an oracle rests on.

First, the authority behind LLM-based oracles often does not come from a specification at all. That makes them useful on everyday software, which rarely has a written specification. It also makes them risky: an oracle resting only on the model's own knowledge, with no external anchor at all, can be wrong with no signal that it is.

Second, the source of authority and the way pass or fail is computed are separate choices (Fig.~\ref{fig:alluvial}). A label such as ``LLM-as-judge'' therefore says nothing about how far to trust the verdict.

Third, the most frequently reported risks concern oracle strength: the most common reported \emph{correctness} failure is an oracle too weak to catch the bug it should. This compounds the hazard studies engage with most, the model's tendency to hallucinate behavior the system never shows.

\subsection{Choosing a source of authority in practice}
The taxonomy also reads as a practical guide: the right source depends on what the project already has.

If the code has documentation, such as docstrings or requirements, specification-derived oracles are the best-supported path. The reviewed work most often turns natural-language documentation into assertions that run against the program~\cite{p1131,p0259,p0698}.

If there is no specification but a trusted second implementation exists, a reference-differential oracle is natural: run both and compare. A newer variant has the LLM write several implementations and treats their agreement as the expected answer, removing the need for an existing reference~\cite{p0699,p1985}.

If the program is interactive or visual, such as a mobile app or a \revI{GUI}, the field uses model-parametric oracles. The LLM judges the output directly from its own knowledge, because no executable specification is available~\cite{p0372,p0417}. This is the least anchored option, so it pairs best with human review or independent cross-checks.

If existing tests already encode expected behavior, implementation-derived oracles can extend them. They inherit the limits of those tests, however, and cannot catch a bug already in the current code.

In every case the source should be stated plainly.

How the model obtains that authority does not change which source it is. A study that uses \revI{retrieval-augmented generation (RAG)} or an agentic tool call to fetch a specification is still specification-derived; one that fetches or executes existing code is still implementation-derived. \revI{Twenty-four of the 83 studies use RAG or agentic (multi-agent) adaptation (14 and 13 respectively, with 3 studies using both). They span six sources (12 specification-derived, 5 implementation-derived, 3 model-parametric, 2 human-elicited, 1 reference-differential, 1 regression-from-prior-version)}, \revI{indicating} that adaptation strategy and source of authority are \revI{separate, cross-cutting} choices.

\subsection{Gaps and a research agenda (\revH{RQ3.c})}
\label{sec:gaps}
Reading the sparse and empty regions of the source-by-mechanism space (Fig.~\ref{fig:alluvial}) and the source totals (Fig.~\ref{fig:source}) yields a concrete agenda. \revI{The gaps below are an interpretive reading of where LLM-based oracles have not yet been applied, not an observed empirical result; several are shaped by our review scope (Section\mbox{~\ref{sec:threats}}) rather than by the state of practice.}

\begin{description}[style=unboxed, leftmargin=0pt, itemsep=5pt]
\item[\revI{Near-empty source categories.}] \revI{One of the seven sources remains unoccupied, and one nearly so. No included study rests on a purely implicit-intrinsic oracle (crashes or well-formedness) as the LLM's contribution: that failure signal is supplied without an LLM, so such designs fall to E2, and the category surfaces only as a \emph{secondary}, compositional signal (in four studies). This cell is therefore empty partly by construction of our scope. Regression-from-prior-version authority, unoccupied in the database-search corpus, is now held by two studies in which a prior version's behavior anchors an LLM-generated or LLM-judged catching test\mbox{~\cite{p2260,p2271}}; with two studies it is the sparsest occupied source. Whether an LLM can add value in the implicit-intrinsic cell, where it is currently bypassed, this corpus cannot settle.}
\item[Empty source-by-mechanism cells.] \revI{LLM-as-judge is never applied to reference-differential authority, and touches implementation-derived authority only once, as a validation step over an inferred contract\mbox{~\cite{p2253}}: when the authority is the code or an independent reference, the included studies otherwise execute or match rather than asking the model to judge.} Whether an LLM judge briefed with the implementation or the reference could outperform exact matching, particularly for outputs that are semantically equivalent but textually different, is \revI{almost} untested. \revI{Human-elicited authority, sparse in the database-search corpus, now covers seven studies spanning five mechanisms.}
\item[Evaluation gaps.] The reliance on custom benchmarks \revI{(64\%)} and the thin use of mutation analysis to validate oracle fault detection \revI{(13 of 83)} together mean the field lacks a shared, fault-grounded yardstick for oracle quality. A standard oracle-evaluation benchmark scored by mutation killing, rather than by textual similarity to a reference oracle, would be valuable.
\item[Trust gaps.] Hallucination is the hazard studies engage with most \revI{(60 of 83)}, but it is named far more often than measured: \revI{58 studies flag it as a risk while 44 report observing it. The 46 mitigations}, in turn, are varied and rarely evaluated for residual risk. The weak-assertion failure mode \revI{(38)} shows that generated oracles frequently under-constrain behavior in a way that similarity metrics do not detect. Measuring oracle \emph{strength}, beyond oracle-generation accuracy, is the open problem these findings point to.
\end{description}

\section{Related Reviews}
\label{sec:related}
Fourteen secondary studies overlap this review's scope~\cite{r288,r310,r322,r1815,rW1,rH1,r141,r273,r421,r1697,r1929,r349,r500,r1838}. The full per-study analysis is given in Table~\ref{tab:reviews}. None organizes the oracle literature by the source of verdict authority. Where they address the oracle at all, thirteen treat it by output form or by LLM technique, and one partitions it by mechanism~\cite{r141}. The only oracle-dedicated secondary study is a non-systematic roadmap~\cite{r310}. None reports an inter-reviewer agreement coefficient, even among the broad reviews of LLMs for software engineering~\cite{rH1,rW1}. Nor do they separate the three questions this review keeps apart: what shape the oracle takes, how pass or fail is computed, and where its authority comes from. Among the surveys we identified, none is simultaneously oracle-focused, systematically searched, organized by source of authority, and reliability-reported; that is the position this review occupies.

\begin{table}[!ht]
\centering
\caption{This review against the 14 related secondary studies (counts over the 14).}
\label{tab:reviews}
\footnotesize
\begin{tabular}{@{}p{5.4cm} c c@{}}
\toprule
Property & Prior reviews & This \\
 & (of 14) & review \\
\midrule
Oracle is the primary unit of analysis & 1\textsuperscript{$\dagger$} & yes \\
Systematic search with a reported funnel & 7 & yes \\
Organized by source of verdict authority & 0 & yes \\
Inter-reviewer agreement ($\kappa$) reported & 0 & yes \\
\bottomrule
\end{tabular}

\smallskip
{\footnotesize \textsuperscript{$\dagger$}The one oracle-focused secondary study is a non-systematic roadmap (no search funnel). One further review partitions oracles by mechanism, counted as \emph{not} organized by source of authority.}
\end{table}

\section{Threats to Validity}
\label{sec:threats}
\subsection{\revH{Search and screening validity}}
The recall-tuned query admitted many off-topic records. These were removed by criterion-based screening (Table~\ref{tab:kappa}).

The pre-filter's exclusions carry the only residual risk: the model-excluded records not individually re-screened. The audited sample puts the false-omission rate at 0.48\% (1/207\revI{, Clopper--Pearson exact 95\% CI 0.01--2.66\%}), projecting to roughly nine \revI{(95\% CI $\approx$0--50)} missed records across the 1{,}860 unaudited exclusions. Because the CI around a single observed omission is wide, this residual risk is disclosed here, with the sampled decisions released for audit. It is unlikely to change the main taxonomy-level findings\revI{, and its consequence can be bounded. A missed record would still have had to clear full-text eligibility, which 54 of the 114 unique full-text records did, so the nine projected omissions correspond to roughly four further included studies. Distributed like the rest of the corpus, they would leave every source ranking unchanged. The finding most sensitive to them is the narrow majority that reaches a verdict without a specification (42 of 83): a worst case in which all four additions were specification-derived would turn that majority into a narrow minority, though the substantive point, that a large share of these oracles need no specification at all, holds either way}.

\subsection{\revH{Scope validity}}
The search covered Scopus, IEEE~Xplore, and the ACM Digital Library; Web of Science, Springer Link, and dedicated preprint servers were not queried, and preprint-only work is therefore under-represented (criterion E5). \revI{The citation-searching round partially mitigates this boundary, reaching work indexed outside the three databases through the citation graph of the included studies.}

LLM-based assertion generation for hardware (criterion E3) is a sizable adjacent literature that we deliberately place out of scope. It is a natural target for a companion review.

\subsection{\revH{Classification validity}}
The source-of-authority code is the review's central interpretive judgment and is, like any qualitative coding, subject to reasonable disagreement at category boundaries. To make it auditable, every included and excluded record is released\revI{~\mbox{\cite{zenodo}}} with its eligibility decision, the criterion applied, the primary and secondary source codes, and an anchoring quotation. The load-bearing boundary rules, notably runtime execution versus static proof, the LLM as oracle versus as test-input generator, \revI{and a supplied specification versus a bare task prompt (Section\mbox{~\ref{sec:framework}}),} are stated explicitly so that any classification can be re-traced and contested. \revI{The source-of-authority coding (Section~\mbox{\ref{sec:codebook}}) is interpretive, so a misclassification could pass undetected; the anchoring quotations in the released sheet are the audit trail for exactly this risk. As a sensitivity check, reassigning the two contested primary-source codings\mbox{~\cite{p0184,p1850}} to their alternative readings changes no aggregate count by more than one study and no conclusion of the review.}

\section{Conclusion}
Test oracles decide whether software behaves correctly, and LLMs have made it possible to generate them or to use a model as one. This review organized that new body of work by a question the literature usually skips: where does the verdict's authority come from? \revI{Across 83 studies, the single most common authority is a written description, but just over half need no specification at all.} \revI{The source categories track this split by construction, and the corpus confirms it}\revI{, crossed only by a small partial fringe where the model itself supplies or reads the specification-shaped artifact other designs are handed.} The source of authority cross-cuts how pass or fail is computed, so describing an oracle only by its mechanism leaves out what matters most. The most frequently reported weaknesses are oracles too weak to catch bugs, and verdicts the model can hallucinate. Its evaluation usually checks resemblance to a known oracle, and rarely whether the oracle catches faults. The source-of-authority taxonomy reads this literature on the grounds that set how far a verdict can be trusted. For practitioners, the first question about any LLM oracle should be where its authority comes from, since that, more than the mechanism, is what a pass or fail can ultimately be defended on. \revI{These findings carry the residual uncertainties disclosed in Section~\mbox{\ref{sec:threats}}: a small number of records the automated pre-filter may have missed, databases and preprint venues not searched, and the interpretive nature of qualitative classification.}

\appendices
\section{Per-Study Categorization}
\label{app:papers}
Table~\ref{tab:appendix} lists every included study with its \revI{primary and, where compositional, secondary source of authority, its specification-dependence label, all coded oracle forms and adjudication mechanisms}, and application domain\revI{, so the multi-valued totals of Tables\mbox{~\ref{tab:formmech}} and\mbox{~\ref{tab:srcmech}} can be reconstructed row by row}. It lets any aggregate count in the synthesis be traced back to the studies behind it. The complete coding sheet, with its coded fields and the anchoring quotations, is provided as supplementary material.

\onecolumn
{\footnotesize
\setlength{\tabcolsep}{3.5pt}
\begin{longtable}{@{}l l l l l l l l@{}}
\caption{The \revI{83} included studies and their complete coding: primary source, \revI{secondary source where a system is compositional, specification dependence (Sp), and all applicable oracle forms and adjudication mechanisms, so every multi-valued aggregate in the synthesis (Tables~\ref{tab:formmech} and~\ref{tab:srcmech}) can be reconstructed from this table. Column abbreviations are defined in the note beneath the table.}}\label{tab:appendix}\\
\toprule
ID & Yr & Source & \revI{2nd src} & \revI{Sp} & \revI{Form(s)} & \revI{Mechanism(s)} & Domain \\
\midrule
\endfirsthead
\multicolumn{8}{@{}l}{\footnotesize\emph{Table~\ref{tab:appendix} continued}}\\
\toprule
ID & Yr & Source & \revI{2nd src} & \revI{Sp} & \revI{Form(s)} & \revI{Mechanism(s)} & Domain \\
\midrule
\endhead
\midrule
\multicolumn{8}{r@{}}{\footnotesize continued on next page}\\
\endfoot
\bottomrule
\endlastfoot
0001\,\cite{p0001} & 2025 & Spec & -- & D & Assert; Exp-out & Exact & web \\
0009\,\cite{p0009} & 2025 & Spec & Human & D & Assert; Property & Human-rev; RT-exec & general-SW \\
0014\,\cite{p0014} & 2025 & Spec & Human & D & Metam & LLM-judge; Human-rev & scientific/numeric \\
0017\,\cite{p0017} & 2025 & Human & Ref-diff & I & Judge & LLM-judge & general-SW \\
0018\,\cite{p0018} & 2026 & Ref-diff & Model & I & Exp-out; Metam & Exact; Statist & ML/DL-system \\
0041\,\cite{p0041} & 2024 & Model & Human & I & Judge & LLM-judge & mobile \\
0043\,\cite{p0043} & 2025 & Impl & Human & I & Assert; Exp-out & Exact; Statist & general-SW \\
0048\,\cite{p0048} & 2026 & Impl & -- & I & Invar & RT-exec & general-SW \\
0072\,\cite{p0072} & 2026 & Model & Regression & I & Assert & Exact; Statist & ML/DL-system \\
0083\,\cite{p0083} & 2026 & Spec & Impl & P & Exp-out & Exact & embedded/PLC \\
0087\,\cite{p0087} & 2025 & Spec & Impl & D & Invar & LLM-judge & API/REST \\
0091\,\cite{p0091} & 2024 & Spec & Model & P & Assert; Exp-out & LLM-judge; RT-exec & general-SW \\
0098\,\cite{p0098} & 2025 & Impl & Regression & I & Assert & Exact; RT-exec & general-SW \\
0103\,\cite{p0103} & 2025 & Human & Ref-diff & I & Judge & LLM-judge; Statist & general-SW \\
0129\,\cite{p0129} & 2025 & Impl & Model & I & Assert & Exact & general-SW \\
0132\,\cite{p0132} & 2025 & Impl & -- & I & Assert & RT-exec & general-SW \\
0137\,\cite{p0137} & 2025 & Spec & -- & D & Assert; Property & RT-exec & API/REST \\
0161\,\cite{p0161} & 2024 & Impl & Human & I & Assert & Exact; RT-exec; Semantic & general-SW \\
0177\,\cite{p0177} & 2026 & Spec & Impl & D & Assert; Property & RT-exec & embedded/PLC \\
0183\,\cite{p0183} & 2024 & Spec & -- & D & Assert & RT-exec & general-SW \\
0184\,\cite{p0184} & 2025 & Model & Impl & I & Assert & Human-rev; RT-exec & general-SW \\
0211\,\cite{p0211} & 2024 & Spec & -- & D & Metam & RT-exec & web \\
0213\,\cite{p0213} & 2025 & Spec & Impl & D & Assert; Metam & LLM-judge; Statist & general-SW \\
0223\,\cite{p0223} & 2025 & Spec & Human & D & Assert & Exact; RT-exec & general-SW \\
0231\,\cite{p0231} & 2025 & Model & Implicit & I & Metam & Exact & embedded/PLC \\
0248\,\cite{p0248} & 2025 & Spec & Model & P & Metam & Human-rev; RT-exec & general-SW \\
0259\,\cite{p0259} & 2025 & Spec & Model & P & Assert; Other & Exact; RT-exec & general-SW \\
0261\,\cite{p0261} & 2026 & Spec & Impl & I & Metam & RT-exec & general-SW \\
0279\,\cite{p0279} & 2025 & Spec & Model & D & Assert; Property & RT-exec & general-SW \\
0287\,\cite{p0287} & 2025 & Model & Human & I & Judge & LLM-judge & GUI \\
0304\,\cite{p0304} & 2025 & Spec & -- & D & Assert; Exp-out & RT-exec & general-SW \\
0316\,\cite{p0316} & 2023 & Model & Human & I & Metam & Human-rev & ADS/autonomous \\
0320\,\cite{p0320} & 2025 & Spec & Impl & D & Assert & RT-exec & mobile \\
0323\,\cite{p0323} & 2025 & Spec & Implicit & P & Metam & Human-rev; RT-exec & ADS/autonomous \\
0339\,\cite{p0339} & 2025 & Spec & -- & D & Assert; Exp-out & RT-exec & general-SW \\
0371\,\cite{p0371} & 2025 & Impl & Regression & I & Assert & Exact; RT-exec & general-SW \\
0372\,\cite{p0372} & 2025 & Model & Human & I & Judge & LLM-judge & mobile \\
0388\,\cite{p0388} & 2025 & Impl & Model & I & Assert & RT-exec & general-SW \\
0390\,\cite{p0390} & 2026 & Spec & -- & D & Judge & LLM-judge & web \\
0406\,\cite{p0406} & 2025 & Impl & Spec & I & Assert; Invar; Property & Formal; RT-exec & security \\
0408\,\cite{p0408} & 2025 & Spec & Impl & D & Assert; Invar; Property & Human-rev; RT-exec & security \\
0413\,\cite{p0413} & 2025 & Impl & -- & I & Assert & Exact & general-SW \\
0417\,\cite{p0417} & 2025 & Model & Human & I & Judge & LLM-judge & GUI \\
0464\,\cite{p0464} & 2026 & Model & Implicit & I & Judge; Metam & LLM-judge; Semantic & mobile \\
0698\,\cite{p0698} & 2024 & Spec & -- & D & Assert & RT-exec & general-SW \\
0699\,\cite{p0699} & 2023 & Ref-diff & Model & I & Exp-out & Exact; RT-exec & general-SW \\
1131\,\cite{p1131} & 2025 & Spec & Impl & P & Assert; Exp-out & Exact; RT-exec & general-SW \\
1154\,\cite{p1154} & 2025 & Impl & Regression & I & Assert; Exp-out & Exact; RT-exec & general-SW \\
1311\,\cite{p1311} & 2025 & Spec & -- & D & Exp-out; Property & Exact; RT-exec & embedded/PLC \\
1725\,\cite{p1725} & 2025 & Spec & Human & D & Assert; Judge & LLM-judge; RT-exec & web \\
1846\,\cite{p1846} & 2026 & Spec & Impl & D & Assert; Exp-out & LLM-judge; RT-exec & general-SW \\
1850\,\cite{p1850} & 2026 & Ref-diff & -- & I & Metam & Exact; Formal; RT-exec & other \\
1985\,\cite{p1985} & 2026 & Ref-diff & Spec & P & Assert; Exp-out & RT-exec; Semantic & general-SW \\
2138\,\cite{p2138} & 2026 & Spec & Ref-diff & P & Judge & LLM-judge; Semantic & general-SW \\
2246\,\cite{p2246} & 2022 & Model & Human & I & Assert; Exp-out & LLM-judge; RT-exec & general-SW \\
2247\,\cite{p2247} & 2023 & Model & Spec & P & Assert; Judge & Exact; RT-exec & general-SW \\
2248\,\cite{p2248} & 2023 & Model & Human & P & Assert; Exp-out; Other & RT-exec & general-SW \\
2249\,\cite{p2249} & 2026 & Impl & -- & I & Assert; Metam & Exact; RT-exec; Statist & general-SW \\
2250\,\cite{p2250} & 2026 & Impl & -- & I & Assert & Other; RT-exec & general-SW \\
2251\,\cite{p2251} & 2024 & Human & -- & I & Assert & Exact; RT-exec & general-SW \\
2252\,\cite{p2252} & 2026 & Human & Model & I & Assert; Exp-out & LLM-judge; Exact; Human-rev; RT-exec & embedded/PLC \\
2253\,\cite{p2253} & 2026 & Impl & Model & I & Assert; Invar & LLM-judge; RT-exec & general-SW \\
2254\,\cite{p2254} & 2026 & Spec & Implicit & D & Assert; Invar & RT-exec & web \\
2255\,\cite{p2255} & 2025 & Spec & Impl & P & Assert; Property & RT-exec & general-SW \\
2256\,\cite{p2256} & 2025 & Model & Human & I & Judge & LLM-judge & mobile \\
2257\,\cite{p2257} & 2024 & Spec & Impl & D & Assert; Invar & Formal & general-SW \\
2258\,\cite{p2258} & 2026 & Model & Spec & P & Assert; Exp-out & LLM-judge; Exact; RT-exec; Statist & general-SW \\
2259\,\cite{p2259} & 2026 & Spec & -- & D & Assert & Other; RT-exec & general-SW \\
2260\,\cite{p2260} & 2026 & Regression & Human & P & Assert; Judge & LLM-judge; Human-rev; Other; RT-exec & general-SW \\
2261\,\cite{p2261} & 2026 & Impl & Human & I & Assert & RT-exec & general-SW \\
2262\,\cite{p2262} & 2026 & Impl & -- & I & Assert & RT-exec & general-SW \\
2263\,\cite{p2263} & 2025 & Model & Impl & I & Assert; Judge & LLM-judge & general-SW \\
2264\,\cite{p2264} & 2025 & Human & -- & I & Assert & RT-exec & general-SW \\
2265\,\cite{p2265} & 2025 & Human & Regression & I & Assert & Exact; RT-exec & general-SW \\
2266\,\cite{p2266} & 2026 & Spec & -- & D & Assert; Invar & Formal; Human-rev & embedded/PLC \\
2267\,\cite{p2267} & 2025 & Spec & Human & D & Invar & Formal & general-SW \\
2268\,\cite{p2268} & 2025 & Human & Spec & P & Assert & RT-exec & general-SW \\
2269\,\cite{p2269} & 2025 & Spec & -- & D & Assert; Invar & Formal; Human-rev; RT-exec & security \\
2270\,\cite{p2270} & 2026 & Model & Impl & I & Assert; Judge & Statist & general-SW \\
2271\,\cite{p2271} & 2024 & Regression & -- & I & Assert & RT-exec & general-SW \\
2272\,\cite{p2272} & 2025 & Ref-diff & Spec & P & Exp-out & Exact; RT-exec & general-SW \\
2273\,\cite{p2273} & 2024 & Model & Human & I & Assert; Exp-out & Exact; RT-exec & general-SW \\
2274\,\cite{p2274} & 2023 & Impl & Spec & I & Invar; Property & RT-exec & general-SW \\
\end{longtable}

\vspace{1pt}
\noindent\revI{\textit{Abbreviations.} \textbf{Source of authority:} Spec, specification-derived; Impl, implementation-derived; Model, model-parametric; Ref-diff, reference-differential; Human, human-elicited; Regression, regression-from-prior-version; Implicit, implicit-intrinsic (a dash denotes no secondary source). \textbf{Specification dependence (Sp):} D, dependent; P, partly; I, independent. \textbf{Oracle form:} Assert, assertion; Exp-out, expected-output value; Metam, metamorphic relation; Invar, invariant; Property, property-based testing (PBT); Judge, classifier/judge label; Other, exception oracle. \textbf{Adjudication mechanism:} RT-exec, runtime assertion execution; Exact, exact/structural match; LLM-judge, large language model as judge; Human-rev, human review; Statist, statistical/distributional; Semantic, semantic/similarity; Formal, formal verification. \textbf{Domain:} general-SW, general-purpose software; ML/DL, machine/deep learning; PLC, programmable logic controller; API/REST, application programming interface/representational state transfer; ADS, autonomous driving system; GUI, graphical user interface.}
}

\twocolumn

\section*{Data Availability}
The replication package for this review is openly available on Zenodo~\cite{zenodo} under a CC~BY~4.0 license. \revI{Its concept DOI, \mbox{10.5281/zenodo.21194940}, always resolves to the current version.} It contains the review protocol and extraction codebook, the frozen search query and screening log, the complete de-duplicated screening funnel with per-record decisions and reasons, the independent dual-screening decisions, \revI{the Stage-1 screening prompt,} the per-study coding sheet \revI{(with an anchoring quotation for every study)}, \revI{both coding passes for each arm together with the per-record adjudication decisions,} and the PRISMA~2020 flow diagram and checklist. Publisher PDFs, raw database exports, and record abstracts are not redistributed for copyright reasons. The released per-database exports contain bibliographic metadata only.

\section*{Acknowledgment}
The authors thank Noor Fatima (National University of Sciences and Technology, Pakistan) for her help with the final review and corrections of the manuscript. An LLM (Anthropic Claude) was used as the recall-oriented screening pre-filter described in the Methodology, disclosed under PRISMA~2020 item~8. Every retained and sampled screening decision was verified by independent human dual screening.

\begin{IEEEbiography}[{\includegraphics[width=1in,height=1.25in,clip,keepaspectratio]{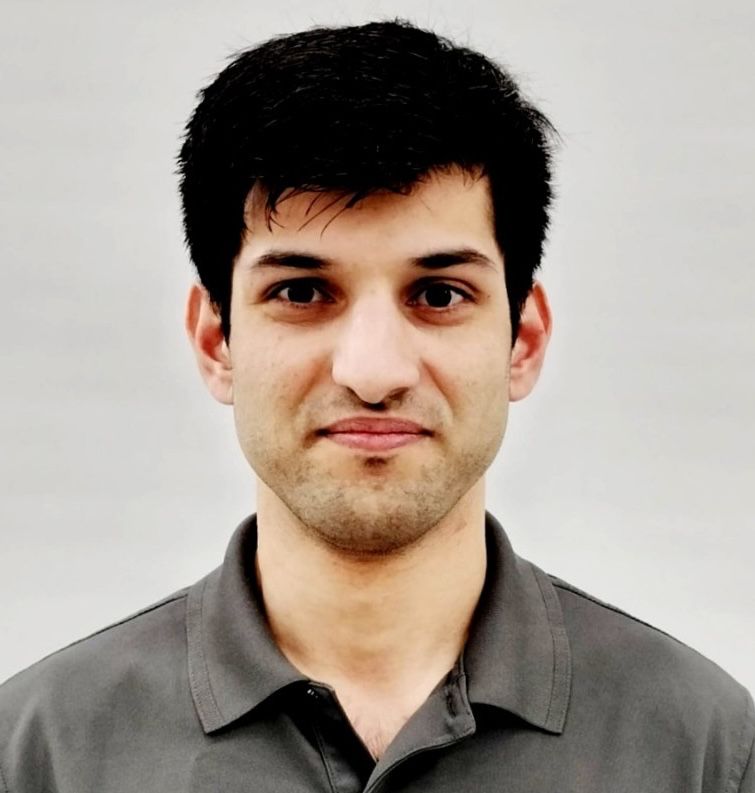}}]{Ali Hassaan Mughal}
\revI{(Graduate Student Member, IEEE)} has worked as a Senior Software Developer and Team Lead at Xpressdocs, and earlier at Paycom and a stealth robotics company. \revI{His work at Xpressdocs centered on software testing and quality assurance, including the design and evaluation of LLM-based multi-agent systems.} He is pursuing an Applied MBA in Data Analytics at Texas Wesleyan University, USA, and holds an M.Sc. in Computer Science from Kansas State University. His research interests include automated software testing, web application quality assurance, applied machine learning, LLMs, and agentic AI systems. \revI{Earlier he was a graduate teaching and research assistant at Kansas State University, where he taught core computer science courses and worked on deep learning for spacecraft trajectory prediction, and a research intern at Aalto University, Finland, on a human computer interaction project. He has published in Elsevier, IEEE, and EAI venues.} \revI{Since 2024 he has served as a reviewer for Expert Systems with Applications, Computers and Electrical Engineering, the International Journal of Information Management, and the ACM Conference on Computer-Supported Cooperative Work and Social Computing (CSCW).}
\end{IEEEbiography}

\begin{IEEEbiography}[{\includegraphics[width=1in,height=1.25in,clip,keepaspectratio]{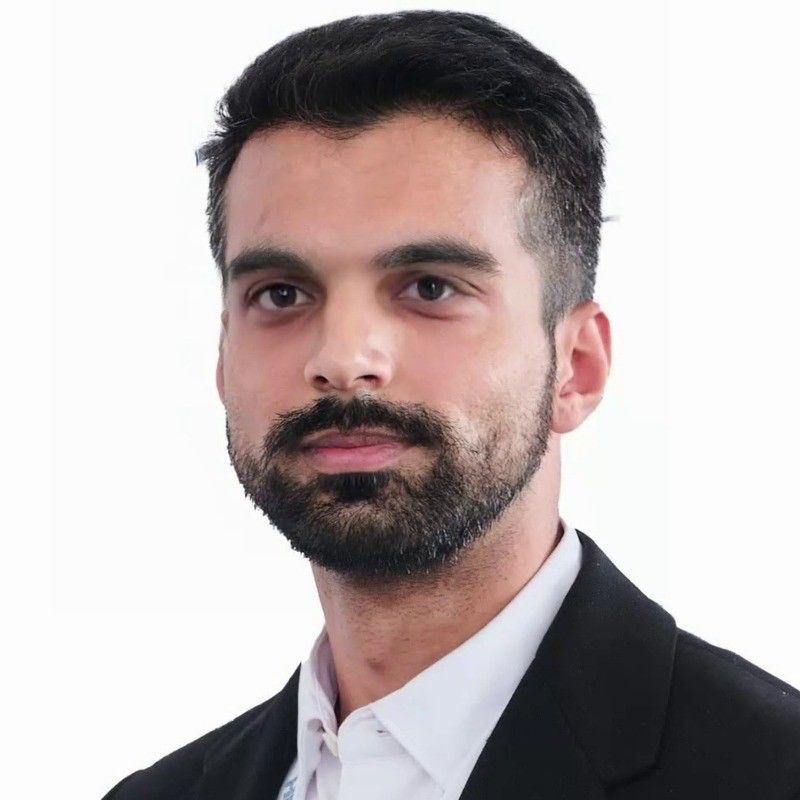}}]{Muhammad Bilal}
is an AI and Digitalization Consultant in the German industrial sector. He holds a Master of Science in Management from the Technical University of Munich, Germany, and has previously worked as a Software Engineer, Business Analyst, and Product Owner. His research interests include the impact of technology on business performance, product quality analytics, the automation of industrial pipelines, LLMs, and agentic AI systems. \revI{His career spans product ownership at Persivia, a clinical AI company; business development at NVision Quantum Technologies, a quantum biotechnology company; healthcare software development at Cognitive Healthcare International; and research at LMU Munich. His current technical work is in agentic systems and LLMs, covering retrieval-augmented generation, agent orchestration, structured output validation, and the evaluation and grounding verification that decide whether such systems can be trusted. He has also published on verifying production LLM applications. His master's thesis and subsequent research examine the sustainability and economics of AI, particularly how its energy use relates to its economic value.} He is the corresponding author.
\end{IEEEbiography}

\vfill

\end{document}